\documentclass[useAMS,usenatbib]{mn2e}
\def\spose#1{\hbox to 0pt{#1\hss}}
\def\approxlt{\mathrel{\spose{\lower 3pt\hbox{$\sim$}}
	\raise 2.0pt\hbox{$<$}}}
\def\approxgt{\mathrel{\spose{\lower 3pt\hbox{$\sim$}}
	\raise 2.0pt\hbox{$>$}}}
\def\approxpropto{\mathrel{\spose{\lower 3pt\hbox{$\sim$}}
	\raise 2.0pt\hbox{$\propto$}}}
\mathchardef\twiddle="2218

\def\multleft#1{\hbox to size{\vbox {\halign {\lft{##}\cr #1}}\hfill}\par}
\def\multright#1{\hbox to size{\vbox {\halign {\rt{##}\cr #1}}\hfill}\par}

\def\today{\ifcase\month\or January\or February\or March\or April\or May\or
      June\or July\or August\or September\or October\or November\or December\fi
      \space\number\day, \number\year}
\def\<{\thinspace}

\def\cm{{\rm\thinspace cm}}

\def\K{{\rm\thinspace K}}
\def\keV{{\rm\thinspace keV}}

\def\km{{\rm\thinspace km}}
\def\kpc{{\rm\thinspace kpc}}

\def\Mpc{{\rm\thinspace Mpc}}

\def\pc{{\rm\thinspace pc}}

\def\s{{\rm\thinspace s}}
\def\yr{{\rm\thinspace yr}}

\usepackage{graphicx}
\usepackage{txfonts}
\usepackage{natbib}
\usepackage{wrapfig}
\usepackage{amssymb}
\usepackage{longtable}
\usepackage{times}
\usepackage[usenames]{color}
\usepackage{psfig}
\usepackage{soul,color}
\usepackage{subfigure}

\usepackage[below]{placeins}
\usepackage{txfonts}
\usepackage{multirow}
\usepackage{array}

\newcommand*{\mysub}[2]{\ensuremath{#1_{\mathrm{#2}}}}

\newcommand*{\msolar}{\mysub{M}{\odot}}
\newcommand*{\zsolar}{\mysub{Z}{\odot}}

\newcommand\ion[2]{#1$\;${\scshape{#2}}}

\title[Stripped Gas Filaments Surrounding M86]
  {Ripping Apart at the Seams: The Network of Stripped Gas Surrounding M86  }
\author[S. Ehlert et al.]
{S.~Ehlert,$^{1,2,3}$\thanks{Email:sehlert@stanford.edu.}
  N.~Werner,$^{1,2,3}$
  A.~Simionescu,$^{1,2,3}$
  S.W.~Allen,$^{1,2,3}$ 
  \newauthor
  J.D.P.~Kenney,$^4$
  E.T.~Million,$^5$
  and A.~Finoguenov$^{6,7}$\\
  $^1$Kavli Institute for Particle Astrophysics and Cosmology, 452 Lomita Mall, Stanford, CA 94305-4085, USA\\
  $^2$SLAC National Accelerator Laboratory, 2575 Sand Hill Road, Menlo Park, CA 94025, USA\\
  $^3$Department of Physiscs, Stanford University, 382 Via Pueblo Mall, Stanford, CA 94305-4060, USA\\
  $^4$Department of Astronomy, Yale University, PO Box 208101, New Haven, CT 06520-8101, USA\\
  $^5$Department of Physics and Astronomy, University of Alabama, 206 Gallalee Hall Box 870324, Tuscaloosa, AL 35487, USA\\
  $^6$Department of Physics, University of Helsinki, Gustaf H\"allstr\"omin katu 2a, FI-00014 Helsinki, Finland\\
  $^7$Center for Space Science Technology, University of Maryland Baltimore County, 1000 Hilltop Circle, Baltimore, MD 21250, USA}

\def\cha{{\it Chandra}}

\def\arcmin {\hbox{$^{\prime}$}}

\def\xmm{{\it XMM-Newton}}
\def\esas{{\it ESAS}}

\begin{document}

\maketitle

\begin{abstract}

\noindent We present a new study of the Virgo Cluster galaxies M86, M84, NGC 4338, and NGC 4438 using a mosaic of five separate pointings with \xmm. Our observations allow for robust measurements of the temperature and metallicity structure of each galaxy along with the entire $\sim 1^{\circ}$ region between these galaxies. When combined with multiwavelength observations, the data suggest that all four of these galaxies are undergoing ram pressure stripping by the Intracluster Medium (ICM). The manner in which the stripped gas trailing the galaxies interacts with the ICM, however, is observably distinct. Consistent with previous observations, M86 is observed to have a long tail of $\sim 1 \keV$ gas trailing to the north-west for distances of $\sim 100-150 \kpc$. However, a new site of $\sim 0.6 \keV$ thermal emission is observed to span to the east of M86 in the direction of the disturbed spiral galaxy NGC 4438. This region is spatially coincident with filaments of H$\alpha$ emission, likely originating in a recent collision between the two galaxies. We also resolve the thermodynamic structure of stripped $\sim 0.6 \keV$ gas to the south of M84, suggesting that this galaxy is undergoing both AGN feedback and ram pressure stripping simultaneously. These four sites of stripped X-ray gas demonstrate that the nature of ram pressure stripping can vary significantly from site to site.

\end{abstract}

\section{Introduction}

Observations of elliptical galaxies in clusters with \xmm \ and \cha \ have shown that the cluster environment has a profound impact on the evolution of galaxies. There are several mechanisms by which galaxies in clusters interact with both neighboring galaxies and the surrounding Intracluster Medium (ICM). Most galaxy-galaxy interaction mechanisms usually require low-velocity approaches to be efficient, and typically result in the stripping of both diffuse interstellar gas and a fraction of their stellar populations. Galaxy-ICM interactions commonly result in the stripping of the gas associated with the galaxy halo. In the most dramatic examples, ram pressure stripping can result in tails of both hot and cool gas trailing behind the galaxies for distances of $\sim 100 \kpc$ or longer.
The physical processes that operate within these tails, however, are not fully understood. While simulations have shown that simple analytic estimates are fairly accurate in predicting the radius of the gas disc that undergoes ram pressure stripping \citep{Gunn1972}, complex physics is required to account for the morphology of the stripped gas tails and their multiwavelength properties \citep[e.g.][]{Roediger2006,Roediger2007,Roediger2008,Tonnesen2010,Tonnesen2011,Tonnesen2012}. The viscosity of the ICM \citep{Roediger2008b}, turbulence, and magnetic fields \citep{Ruszkowski2012} are all expected to influence the morphology of the stripped gas and its evolution after it is deposited into the ICM. The relative contributions of these processes remain uncertain. 

Arguably the ideal site to observe these interaction mechanisms is the nearby, bright Virgo Cluster. The most massive galaxies in this cluster have been studied in detail at multiple wavelengths, and X-ray observations have shown convincing evidence for profound interactions between the galaxies and the ICM. Long tails of cool, stripped gas have been observed trailing in the wake of Virgo ellipticals including NGC 4552 \citep{Machacek2006}, and NGC 4472 \citep{Kraft2011}. 

Observations of the Virgo Cluster have shown that it is composed of three separate sub-groups: one centered on M87, the Brightest Cluster Galaxy (BCG); one centered on M86, an elliptical galaxy roughly one degree ($\sim 300 \kpc$) to the west of M87; and the final sub-group centered on NGC 4472 (also known as M49) approximately $4.5^{\circ}$ to the south of M87 \citep{Binggeli1987,Binggeli1993,Bohringer1994}. Spectroscopic studies of M86 have shown that it has an absolute blue-shifted velocity of $\sim 250 \km \s^{-1}$ while M87 has a red-shifted velocity of $\sim 1300 \km \s^{-1}$. Given an average temperature of the ambient Virgo ICM of $\sim 2.3 \keV$ \citep[][corresponding to a sound speed of $\sim 700 \km \s^{-1}$]{Urban2011}, M86 must be traversing through the Virgo Cluster supersonically with a Mach number of at least $\mathcal{M} \gtrsim 2$. 

In the vicinity of M86, additional galaxies are located within approximately $\sim 100 \kpc$ of one another in projection. The two most massive galaxies, M86 and M84, are separated by a projected distance of only $60 \kpc$. Although these galaxies appear similar in optical imaging studies, further studies utilizing optical spectroscopy and X-ray observations show that they are interacting with the ambient Virgo ICM in different manners. Observations of M86 with \cha \ and \xmm \ \citep{Finoguenov2004,Randall2008} show that it is undergoing significant ram pressure stripping, with diffuse X-ray emitting gas as well as cooler material trailing behind the galaxy for distances of approximately $\sim 150 \kpc$ in projection to the northwest. There is only sparse evidence for ram pressure stripping in observations of M84 with \cha \ \citep{Finoguenov2008}, but there is clear evidence for a powerful AGN outburst at the center of this galaxy \citep{Finoguenov2008}. M84 has an absolute red-shifted velocity of $\sim 1000 \km \s^{-1}$ \citep{Smith2000,Trager2000}, approximately consistent with the measured velocity for M87. The region in between M84 and M86 is also claimed to be the site of bright intracluster light emission \citep{Rudick2010}. Two other galaxies are detected in the X-rays nearby: NGC 4388 and NGC 4438. These spiral galaxies have absolute redshifted velocities of $2566 \km \s^{-1}$ \citep[NGC 4388,][]{Rines2008} and $71 \km \s^{-1}$ \citep[NGC 4438,][]{Kenney1995}, respectively.  

In this work we present, for the first time, an investigation into the thermodynamic structure spanning the entire region across these four galaxies. Utilizing the full information of five separate \xmm \ observations, including new observations of M84, we are able to investigate the temperature and metallicity structure around these galaxies and investigate how interactions of these galaxies with each other and the ICM have transformed their gas reservoirs.

This paper is structured as follows: in Section 2 we discuss the data sets, their initial processing, and the production of images and spectral maps; in Section 3 we present our results; and in Section 4 we interpret their physical implications. We assume a nominal distance of 17 \Mpc \ to the Virgo Cluster, at which 1 arcsecond corresponds to 82 \pc, and 1 arcminute to 4.92 \kpc.

\begin{figure*}
\includegraphics[width=0.95\textwidth]{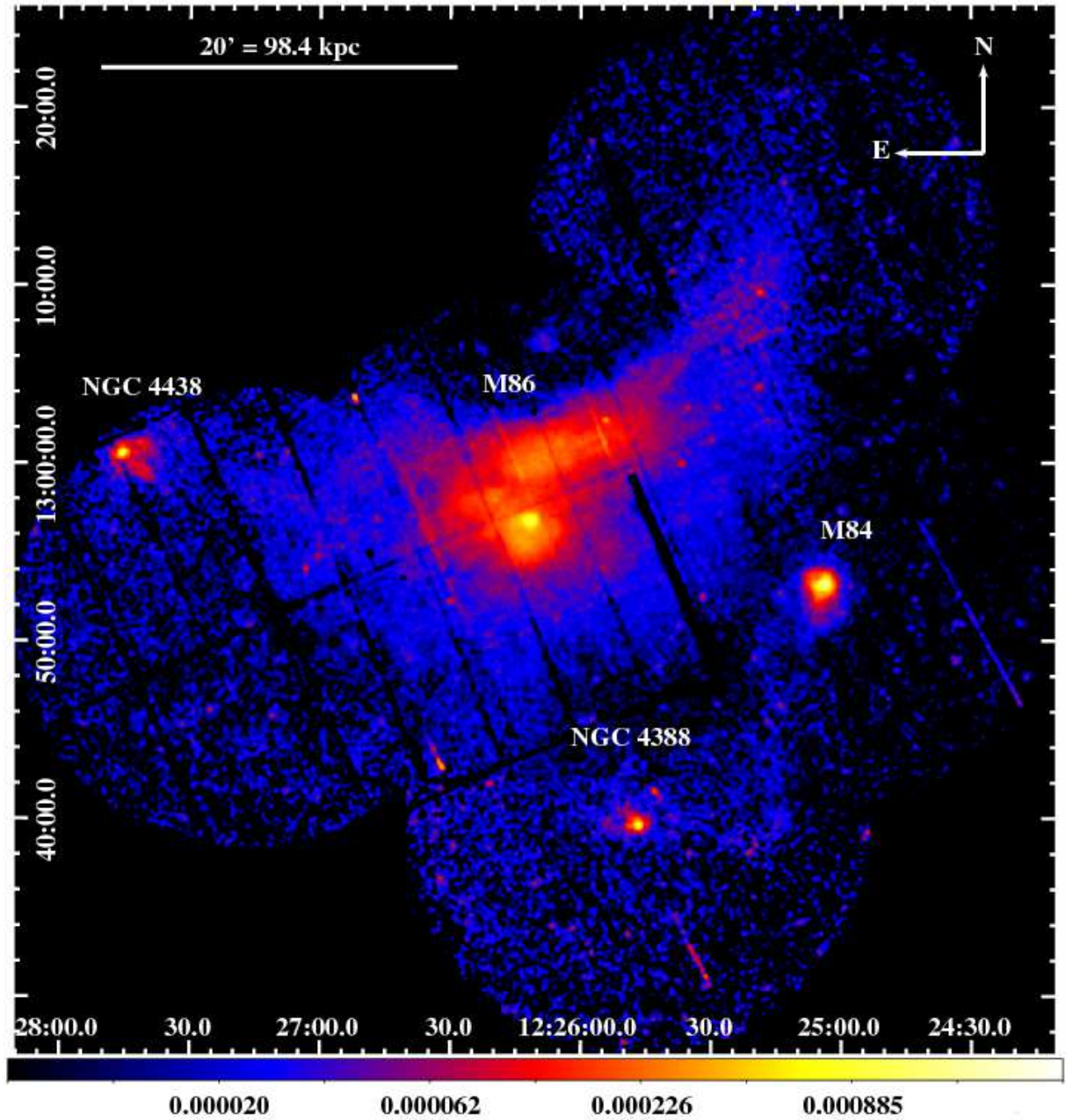}
\caption{\label{Mosaic} Background-subtracted, exposure corrected mosaic of five \xmm \ pointings surrounding M86 (in units of $\rm{cts} \s^{-1}$), utilizing the MOS+PN detectors in the energy band of $0.3-1.0 \keV$. The four X-ray brightest Virgo Cluster members observed in this mosaic (M86, M84, NGC4388, and NGC 4438) are labeled. Diffuse gas is observed trailing to the northwest of M86, to the east of M86 in the direction of NGC 4438, to the south of M84, and to the east of NGC 4388.  }
\end{figure*}

\section{Data Processing and Analysis Methods}\label{datared}
A total of five separate \xmm \ pointings were analyzed, the details of which are listed in Table \ref{Observations}. The data products were cleaned and processed using the \xmm \ Science Analysis System (SAS) (version 11.0), and in particular the Extended Source Analysis Software \citep[\esas,][]{Snowden2008} package included with SAS. We have utilized events from the two MOS and PN detectors for this study. Calibration products and event lists were first produced using the standard SAS routines. Good time intervals (GTI's) for each event list were determined using the {\small MOS-FILTER} and {\small PN-FILTER} routines. Four of the five pointings had little time contaminated by flares, but the most recent observation (of M84, Observation \# 0673310101, PI S. Ehlert) was heavily contaminated by flaring, with only $\sim 40$ ks of the original 130 ks MOS exposure deemed suitable for analysis. No further evidence for flaring was observed in any of the event lists after GTI filtering. 
\begin{table}
\caption{\label{Observations}Summary of the five \xmm \ observations utilized in this study. Exposure times are the net exposure after all cleaning and processing as described in Section \ref{datared} for each detector. The M84 observation (Obs \# 0673310101, denoted by an asterisk) was heavily contaminated by flares. The original event list was split into two separate event lists for each MOS instrument, the summed exposure time of which is shown below.     }
\centering
\begin{tabular}{ c  c c c c} 
\\ \hline {Obs \#} & {Obs. Date} & MOS1 (ks) & MOS2 (ks) & PN (ks) \\ \hline\hline 

0108260201 & July 01 2002 & 62.44 & 63.53 & 43.68\\
0110930701 & December 12 2002 & 8.46 & 10.05  & 3.84\\ 
0210270101 & December 19 2004 & 25.42 & 25.25 & 23.79\\ 
0210270201 & December 27 2004 & 20.38 & 19.92 & 19.02\\ 
0673310101* & June 01 2011 & 37.14 & 42.96 & 15.42\\

\hline
\end{tabular}
\end{table}

\subsection{Imaging Analysis}
Background-subtracted, exposure corrected images of all five pointings were produced in the soft energy band, $0.3-1.0 \keV$. The backgrounds for imaging analysis were determined from customized blank-sky observations \citep{Carter2007} and closed-filter wheel observations to account for the instrumental background. We use blank sky files with similar pointing directions and Galactic absorption column densities to M84 and M86. The closed-filter wheel counts images were subtracted from both the blank-sky and science exposures after renormalizing to match the total counts in each of these images outside of the field-of-view. The remaining blank sky emission was then subtracted from the science images. We also accounted for out-of-time (OOT) events in the PN detector images for each pointing using standard procedures. 

\begin{figure*}
\includegraphics[width=0.95\textwidth]{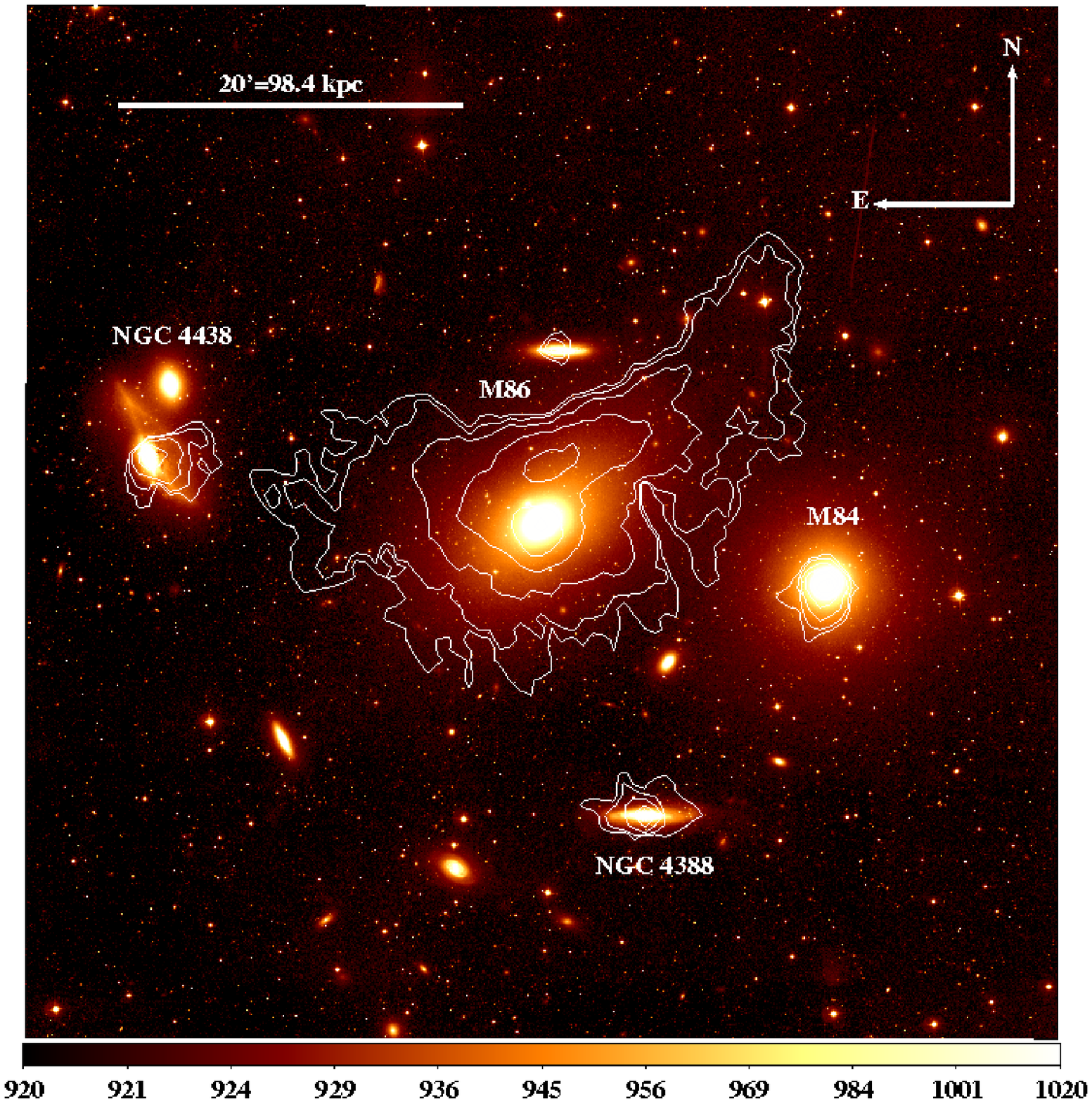}
\caption{\label{Sloan} An r-band image of the region surrounding M86 from the Sloan Digital Sky Survey. The X-ray surface brightness contours are overlaid in white, which show the stripped X-ray emitting gas trailing these galaxies. The stripped gas tail is especially pronounced for the M86 at the center of the image. Tails of stripped gas are observed to the south of M84 and to the northwest of M86. Diffuse soft emission is detected trailing eastward from M86 in the direction of NGC 4438. Faint, diffuse emission is also detected at the position of the edge-on spiral galaxy NGC 4402, located $\sim 10 \arcmin$ to the north of M86.   }
\end{figure*}

\subsection{Thermodynamic Mapping}
Our \xmm \ observations are sufficiently deep for high signal-to-noise spectra to be extracted from relatively small regions. This enables us to carry out detailed, spatially resolved measurements of the thermodynamic quantities across the entire mosaic. All spectral analysis was carried out using {\small XSPEC} \rm  \citep[][version 12.6]{Arnaud2004}. Instrumental background was determined for the source spectra using closed filter wheel observations of the detectors renormalized to have the same count rate in the $10-12 \keV$ energy band. We subtracted the out-of-time events from the PN spectra before performing these fits. We modeled the sky background for each spectral bin using the background model of \cite{Urban2011}, renormalized to account for the appropriate region geometries. This model assumes a sum of three components: an absorbed power law from the unresolved point sources \citep{Deluca2004}, absorbed thermal emission with $kT \sim 0.2 \keV$ from the Galactic halo \citep{Kuntz2000} and the unabsorbed thermal emission from the Local Hot Bubble \citep{Sidher1996,Kuntz2000}. Tests show that no additional background components are necessary.  Allowing the temperatures, photon indexes, or normalizations of the background components to vary did not change the results in a significant manner.

\subsubsection{Regions of Interest}\label{ROI}
We identified the independent regions of interest for our spectral analysis by binning the background subtracted image into regions of equal signal to noise. Here we made use of the Weighted Voronoi Tessellations binning algorithm by \cite{Diehl2006}, which is a generalization of the Voronoi binning algorithm of \cite{Cappellari2003}. We binned the image using two signal-to-noise thresholds (hereafter the low S/N and high S/N maps, respectively). Spectra were extracted from each spectral region, and appropriate response files were generated using the {\small RMFGEN} and {\small ARFGEN} tools. All spectral fits were performed in the energy band $0.4-7.0 \keV$, although we ignored the energy range $1.2-1.8 \keV$ in order to avoid the instrumental Al and Si emission lines in the MOS detectors. In the resulting energy band, there are roughly $\sim 4000$ counts per region in our low S/N maps and $\sim 7000$ counts per region for our high S/N maps.

\subsubsection{Spectral Modeling and Thermodynamic Quantities}

For the low S/N regions, we fit the spectra with an absorbed, single temperature {\small APEC} plasma model. We utilized the photoelectric absorption model of \cite{McCammon1992} and fixed the Galactic absorption column density to the value derived by \cite{Kalberla2005} $\mysub{N}{H}=2.13 \times 10^{20} \cm^{-2}$. We assume the solar element abundance ratios given by \cite{Grevesse1998}. The temperature, metallicity, and the normalization were all independent free parameters. Best fit model parameters were obtained by minimizing the modified C-statistic in {\small XSPEC}. All uncertainties quoted are $68\%$ ($\Delta C=1$) confidence intervals. 

The high S/N spectra were fitted with a similar model, but with an additional thermal component to account for the ambient Virgo Cluster emission viewed in projection. The temperature of this thermal emission was fixed to  2.3 \keV, in agreement with measurements of the diffuse Virgo ICM at the same distance from M87 \citep{Urban2011}. The metallicities of the two thermal components were linked to be the same. The exact choice of the temperature of the ambient Virgo ICM does not have a strong influence on the measured temperature or metallicity for the second component.   

To search for multi-temperature structure in the high S/N map, we also attempted to fit the spectra with a two-temperature plus cooling flow model ({\small APEC+APEC+MKCFLOW}). For the cooling flow component we set the upper and lower temperatures to $1 \keV$ and $0.1 \keV$, respectively. All of our measured temperatures have an uncertainty of less than $\sim 10\%$, regardless of the model.

\section{Results}
The background subtracted, exposure corrected image of the region surrounding M86 is shown in Figure \ref{Mosaic}. To the north-west of M86 a long tail of diffuse gas is observed, which has been discussed in detail by \cite{Finoguenov2004} and \cite{Randall2008}. The length of this tail is $\sim 150 \kpc$ in projection, although given the large line-of-sight velocity of M86 with respect to the ambient Virgo ICM this is likely a conservative lower limit to its true length. The width of this tail in projection, however, is roughly uniform along its entire length. To the east of M86, diffuse emission is observed out to the neighboring galaxy NGC 4438. Diffuse gas is also detected for a distance of approximately $\sim 15 \kpc$ to the south of M84. These tails of X-ray emission trailing M86 and M84 extend beyond the bulk of the optical galaxy emission, as shown in the Sloan Digital Sky Survey image of this field (Figure \ref{Sloan}). There is no evidence for diffuse emission that bridges between M84 and M86. There is also tentative evidence for diffuse X-ray emission trailing to the east of NGC 4388.

\subsection{Temperature Structure}
Maps of the temperature and metallicity structure across the region are shown in Figure \ref{Thermo}. Cool gas with a temperature of $\sim 1 \keV$ is detected in various directions around M86, with the the majority of this gas phase located in the regions immediately surrounding the galaxy and to the north-west of the galaxy. This $\sim 1 \keV$ component traces the morphology of M86's X-ray tail, and originates from the ram pressure stripping of M86 \citep[e.g.][]{Forman1979,Fabian1980,Nulsen1982,White1991,Finoguenov2004,Randall2008}. 

To the east of M86 is a site of cooler X-ray emitting gas, observed to have a temperature of $\sim 0.6 \keV$, spatially coincident with the brighter diffuse X-ray emission spanning between M86 and NGC 4438. Gas with a temperature of $\sim 0.6 \keV$ is also observed just to the north of M86, extending to the northwest in the same direction as the $\sim 1 \keV$ component. Fitting each spectral bin in the region between M86 and NGC 4438 with a two-temperature plus cooling flow model results in no detection of significant cooling flows. For any particular bin, these fits constrain the spectroscopic cooling rate of X-ray gas from $1.0 \keV$ to $0.1 \keV$ to $\lesssim 0.01 \msolar \yr^{-1}$, and the upper limit for the integrated spectroscopic cooling rate across all of the bins between these two galaxies is $\lesssim 0.1 \msolar \yr^{-1}$.  

\begin{figure*}
\includegraphics[width=0.47\hsize]{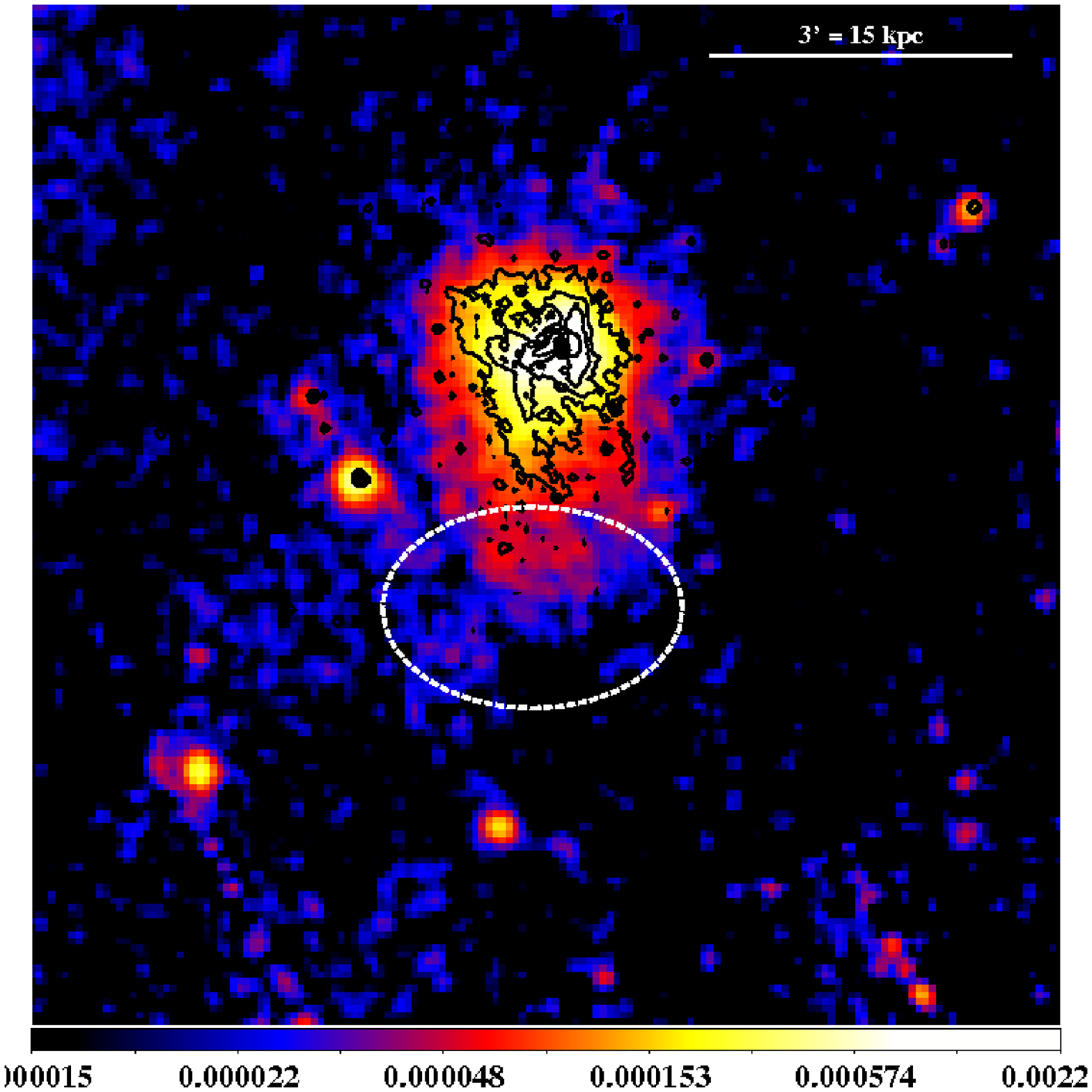}
\includegraphics[width=0.47\hsize]{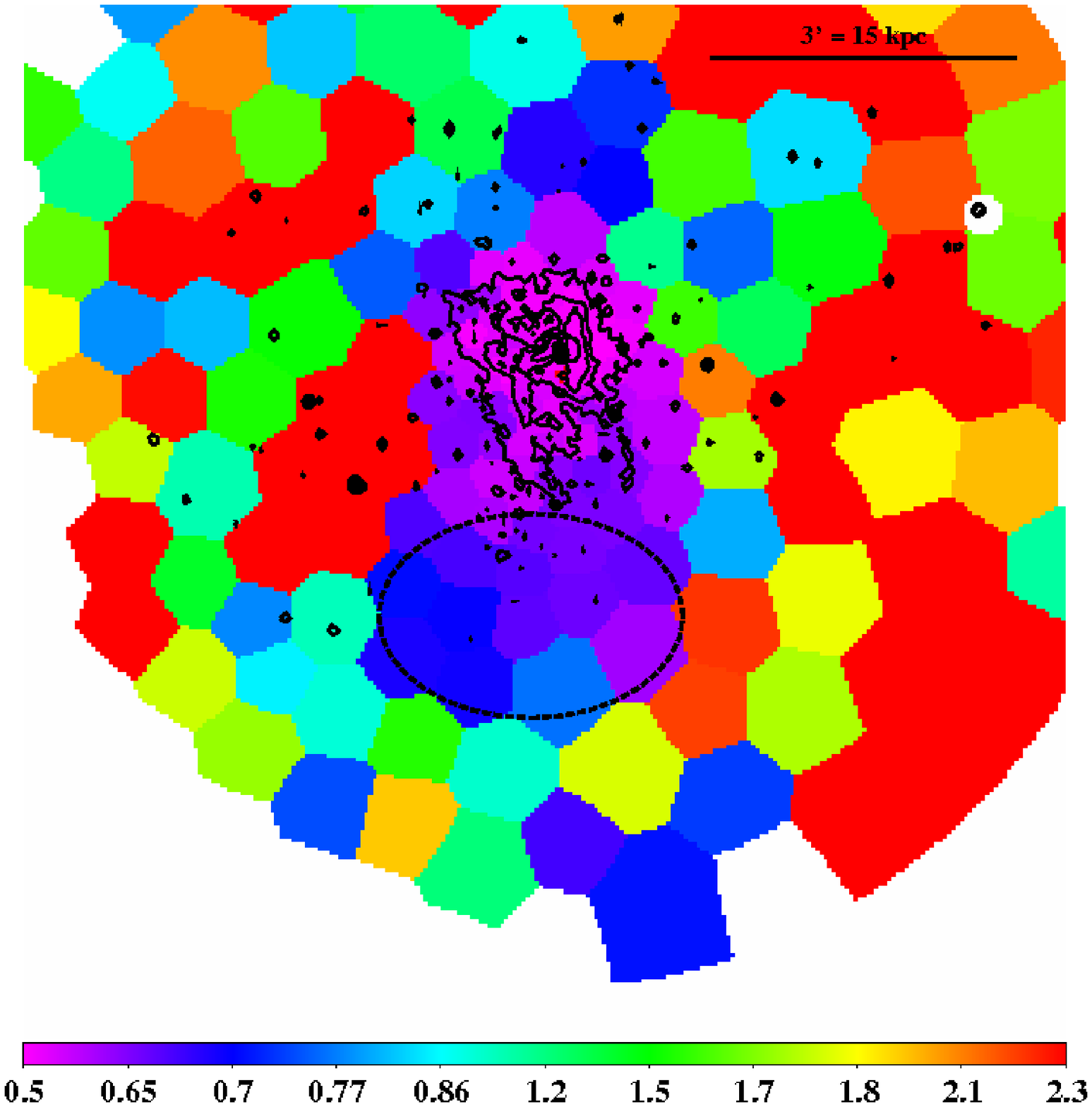}
\caption{\label{M84Zoom} A zoomed in view of M84 with \xmm. Both of these images show the same field of view. The dashed ellipse denotes the region which we use to estimate the rate at which X-ray halo gas is being stripped from M84.  {\it Left:} The background subtracted, exposure corrected image of M84 in the $0.3-1.0 \keV$ energy band. The black contours are derived from \cha \ observations of this galaxy, which show the extent of the AGN-driven cavities \citep{Finoguenov2008}. {\it Right:} The corresponding temperature map of this same region, in units of \keV. The bin map is similar to that described in  \S \ref{ROI}, although with a only $\sim 250$ net counts per bin in order to resolve smaller scale features. Each bin is fit with a single-temperature plasma model with Galactic absorption (\small{PHABS $\times$ APEC}). The same \cha \ contours are overlaid in black. The brighter gas trailing to the south of M84 is shown to be significantly cooler than the ambient ICM.}  
\end{figure*}

In the vicinity of M84, we detect the presence of cool ($\sim 0.7 \keV$) gas trailing to the south of the galaxy, out to a distance of $\sim 15 \kpc$. This is significantly larger than the scale of AGN feedback in this system (see Figure \ref{M84Zoom}). The temperature of this gas is observed to increase smoothly to the south, until it becomes indistinguishable from the ambient ICM. The presence of cool gas trailing to the south of M84 suggests ram pressure stripping of its X-ray halo amidst northward motion. 

We use the presence of this tail to provide an order-of-magnitude estimate of the rate at which X-ray gas is stripped from M84. In order to determine the gas mass in the tail, we assume it has a prolate ellipsoidal geometry with a semi-major axis of $7.5 \kpc$ and a semi-minor axis of $4.2 \kpc$ (shown in Figures \ref{M84Zoom}). This geometry is based on the extent to which we can detect cooler ($\sim 1 \keV$) gas to the south of the cavity regions. We only include the region to the south of the AGN-driven cavities seen in \cha \ observations of M84. Based on the surface brightness enhancement in this region, we estimate that the density of the gas in the tail is $\sim 0.005 \cm^{-3}$, roughly a factor of 2 denser than the surrounding Virgo ICM \citep{Urban2011}. With these assumptions, we determine that a mass of $\sim 7 \times 10^{7} \msolar$ of X-ray gas has been stripped by the bulk motion of M84 through the Virgo ICM. Using the sound crossing time as an estimate for the time scale over which this stripping occurred and an assumed sound speed of $700 \km \s^{-1}$, the overall mass stripping rate is calculated to be $\sim 7 \msolar \yr^{-1}$. Since there is no evidence for jumps in the ICM temperature or surface brightness indicative of a shock front to the north of M84, this galaxy is likely traversing through the ICM at sub-sonic speeds\footnote{Since M84 and M87 have similar line-of-sight velocities, it is also unlikely that M84 has a significant velocity component through the Virgo ICM along the line-of-sight. This maximizes the likelihood that such a shock front, if present, would be detected.}. Resultingly, the mass stripping rate we estimate is likely an upper limit as to the true rate at which M84's X-ray halo gas is being stripped by ram pressure.

To the west of M86 in the direction of M84, a clear temperature gradient is observed over the entire region. Between the two galaxies is a $\sim 20 \kpc$ band of emission consistent with the ambient Virgo ICM, in both one-temperature and two-temperature fits.

\subsection{Metallicity Structure}

\begin{figure*}
\includegraphics[width=0.47\hsize]{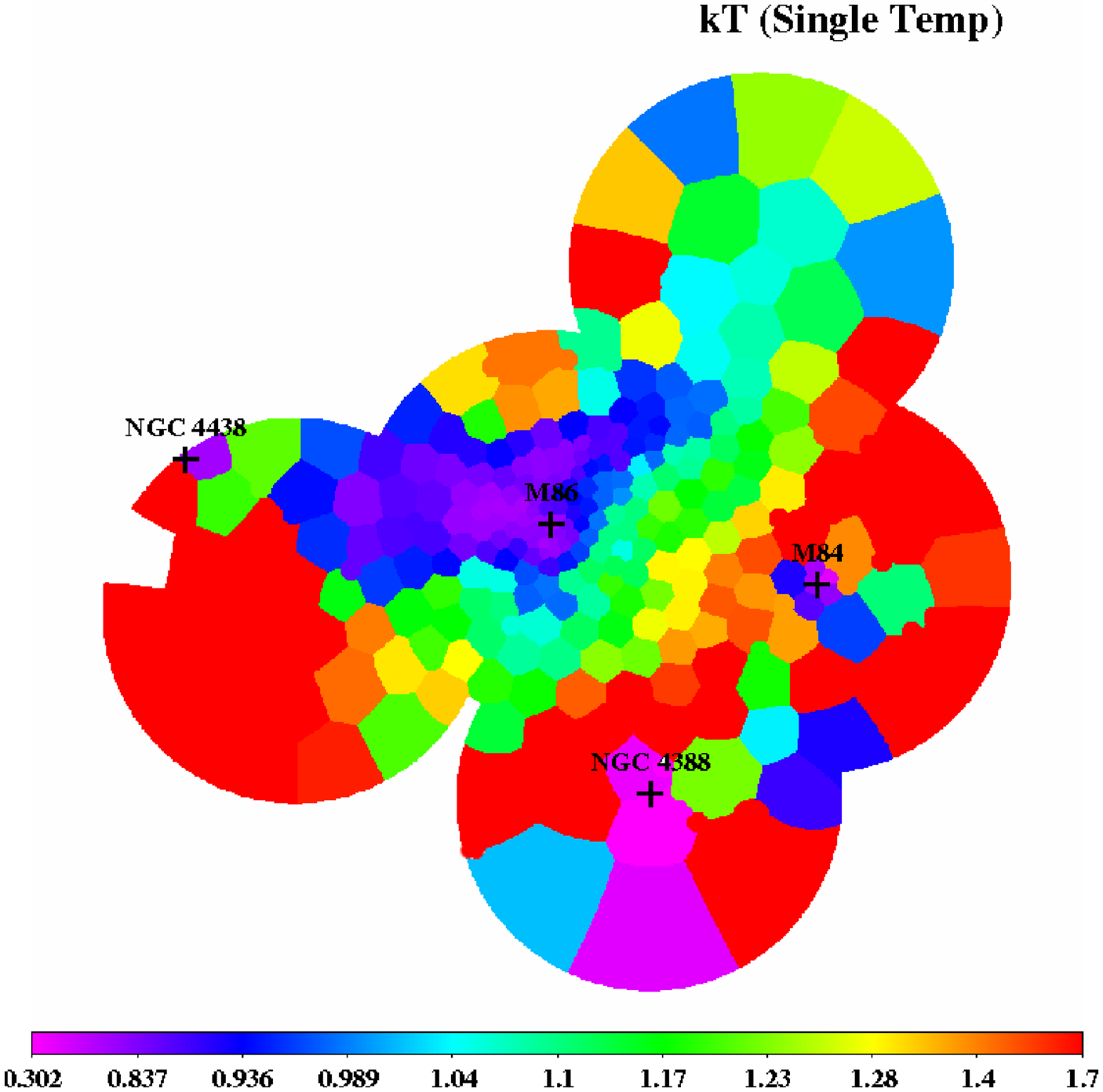}
\includegraphics[width=0.47\hsize]{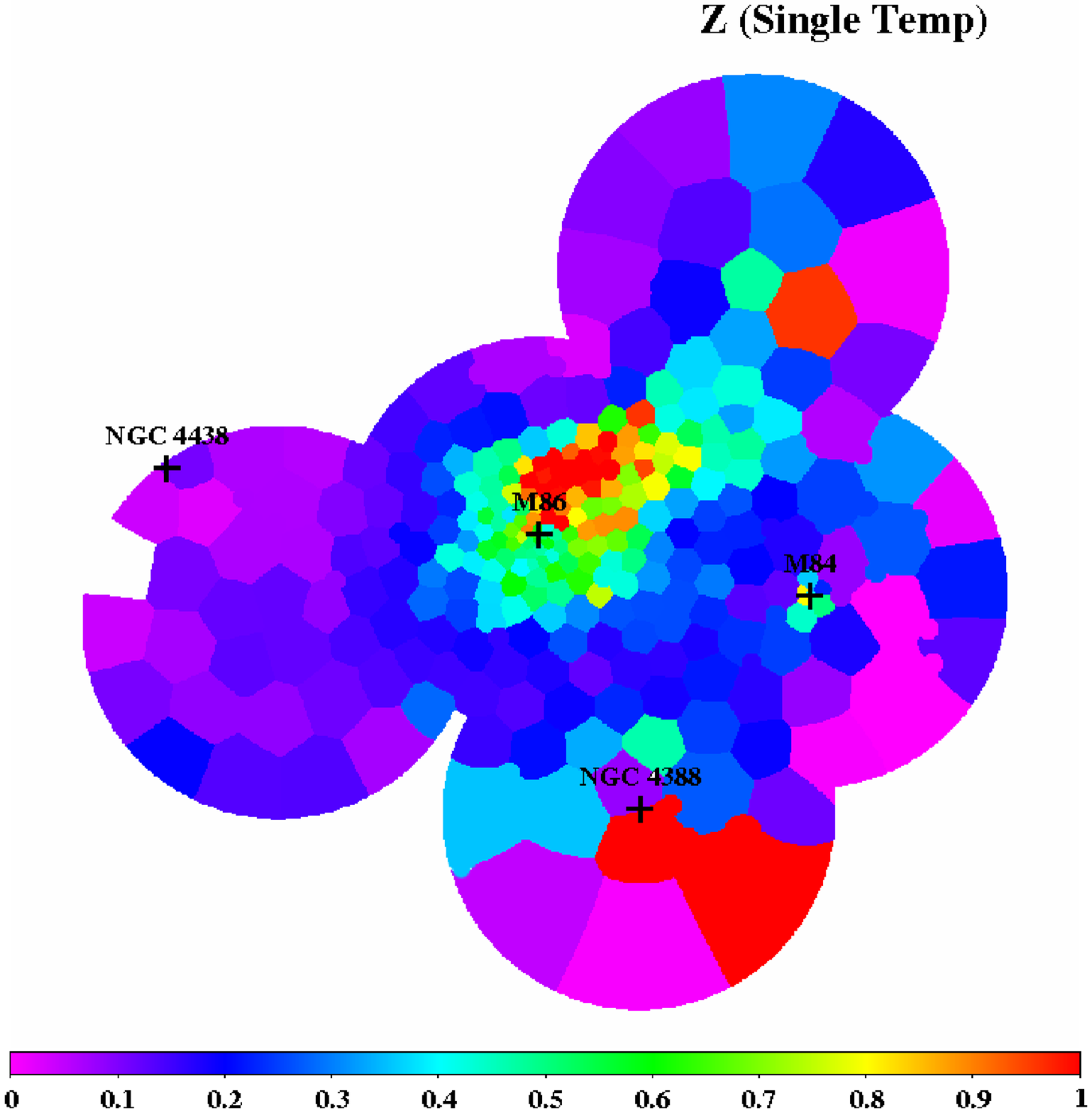}\vspace{0.5mm}
\includegraphics[width=0.47\hsize]{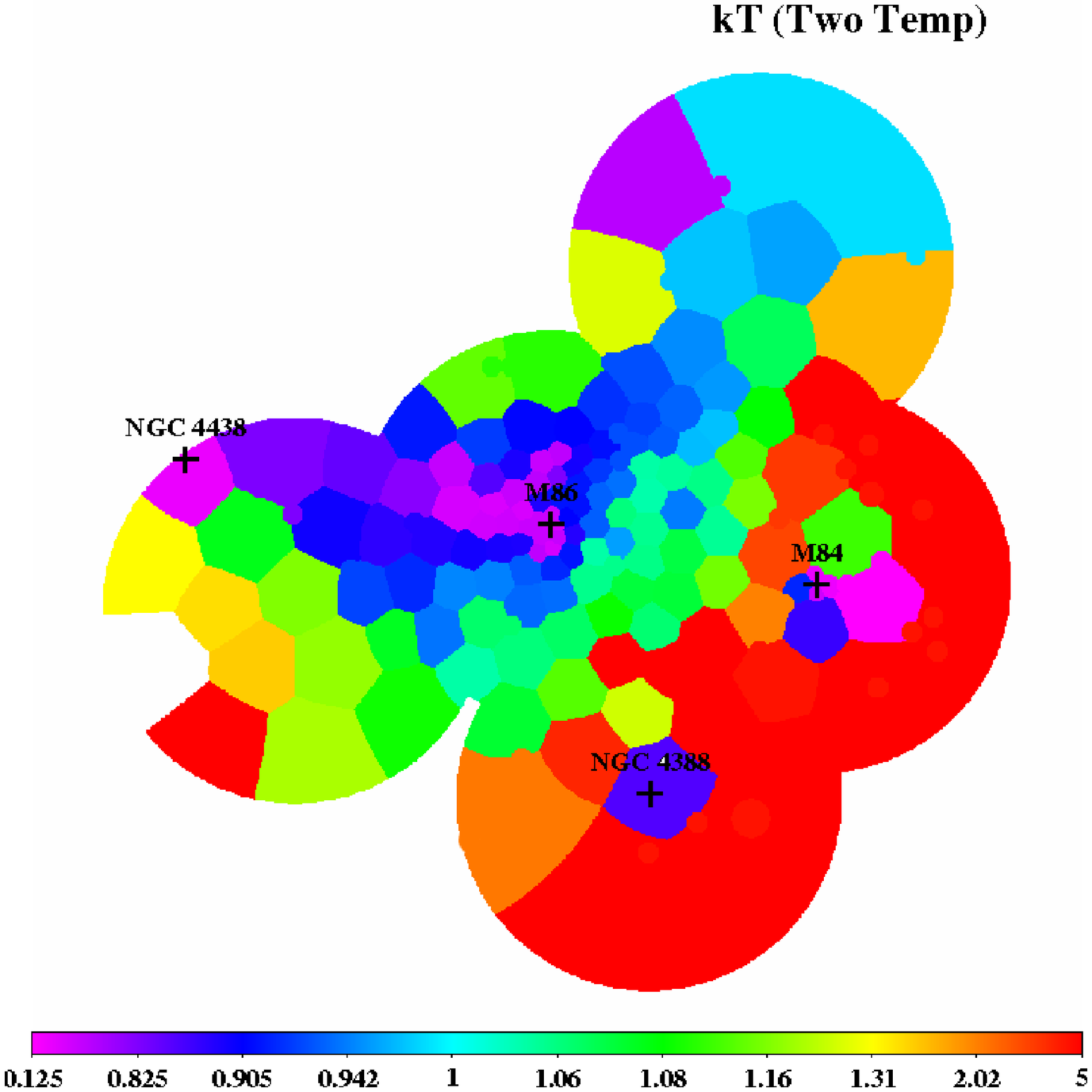}
\includegraphics[width=0.47\hsize]{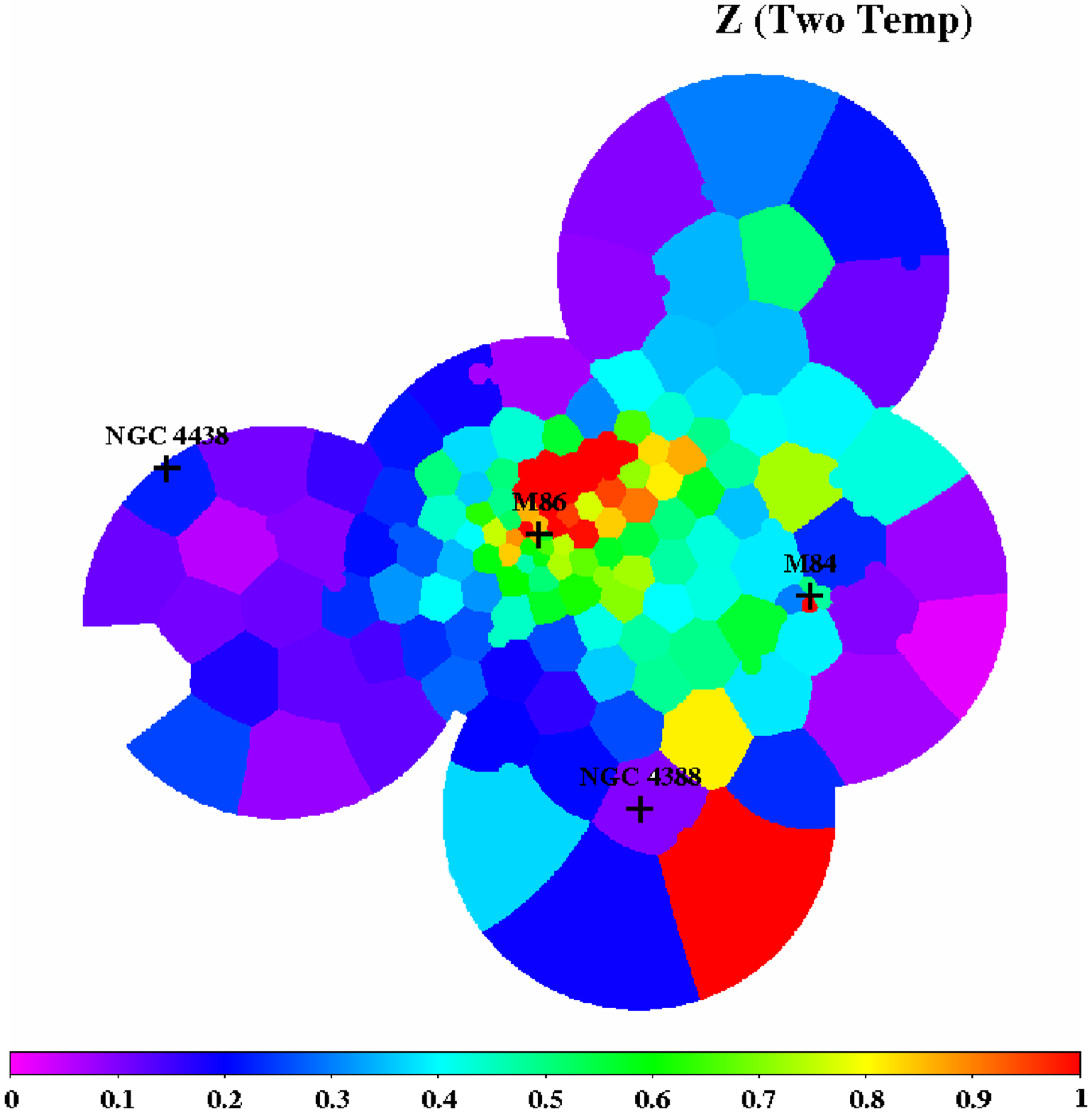}

\caption{\label{Thermo} Temperature and metallicity maps of the $\sim 1^{\circ}$ region surrounding M86, including the galaxies M84, NGC 4388, and NGC 4438. The upper row shows our low S/N ($\sim 4000$ counts per bin) temperature and metallicity maps. The values show our best-fit parameters using a single-temperature thermal emission model. The lower row shows temperature and metallicity maps from our high S/N spectra ($\sim 7000$ counts per bin). The values in these maps are derived from a two-temperature fit, with the second temperature fixed to $2.3 \keV$ and the metallicities of both components tied to the same value. This two-temperature fit optimally identifies regions hosting gas significantly cooler than the ambient Virgo Cluster gas, which has an assumed temperature of $2.3 \keV$. The inclusion of an additional temperature component also reduces metallicity measurement biases.}

\end{figure*}

Although the single-temperature and two-temperature fits qualitatively agree in their metallicity measurements (see Figures \ref{Thermo}b and \ref{Thermo}d), we focus our discussion on the measurements from the two-temperature fits. These measurements are performed at higher signal-to-noise and are less sensitive to biases in metallicity measurements that arise by modeling multi-phase emission with a single temperature \citep{Buote2002}. For comparative purposes, the metallicity of the Virgo cluster at the same radius from M87 \citep{Urban2011} is approximately $Z \sim 0.3 \ \zsolar$. 

The metal-richest gas is located near the center of M86, which has a measured metallicity of at least $\sim 1 \zsolar$. Metal-rich gas ($\gtrsim 0.5 \zsolar$) is observed in the direction of M86's stripped tail, out to distances of $\sim 90-100 \kpc$. There is no strong evidence for metal-rich gas spanning between M86 and NGC 4438. However, it is possible that a metallicity-measurement bias arising from unaccounted temperature phases is still present here. A smaller enrichment in metals ($Z \gtrsim 0.5 \zsolar$) is observed near the core of M84.

\subsection{Ultraviolet Observations with {\it GALEX}}

We have also utilized publicly available ultraviolet (UV) observations of the region between M86 and NGC 4438 with {\it GALEX} in order to search for star formation in this region. We used observation ID ``NGA\_Virgo\_MOS10'' (observation date 2004-03-11), part of the {\it GALEX} Ultraviolet Virgo Cluster Survey \citep{Boselli2011}, which is well centered between the two galaxies. There are no siginficant detections of young stellar populations in the far UV image of this field. Converting the observed surface brightness of far-UV photons into a star formation rate using \cite{Kennicutt1998}, we place a strict upper limit on the star formation rate in the region of the H$\alpha$ filaments of $\lesssim 0.1 \msolar \yr^{-1}$.

\section{Discussion}

Using a mosaic of \xmm \ observations of the region surrounding M86, we have identified multiple sites of stripped X-ray emitting gas associated with all four galaxies: M86, M84, and NGC 4438, and NGC 4388. Although the evidence for stripped X-ray emitting gas near NGC 4388 is weak, radio observations of this galaxy show a tail of \ion{H}{i} gas trailing to the north of this galaxy, with a velocity structure consistent with NGC 4388 \citep[e.g.][]{Oosterloo2005}. All four of these galaxies are therefore undergoing some form of ram pressure stripping, although each galaxy has its own unique observational signatures associated with it. The stripped X-ray emitting gas trailing to the northwest of M86 and the \ion{H}{i} tail trailing to the north of NGC 4388 have been detected and discussed in previous work \citep[e.g.][]{Randall2008,Oosterloo2005}. The stripped X-ray emitting gas to the south of M84 and the cool X-ray emitting gas stretching from M86 to NGC 4438 have not been discussed at length\footnote{The diffuse tail of gas trailing to the south of M84 was discussed briefly in \cite{Randall2008}, although there were insufficient counts in the \cha \ data to resolve its thermodynamic structure.}. We expand our discussion of these two sites of ram pressure stripping below.

While all four of the galaxies lie near one another in projection, the X-ray data combined with absolute velocity measurements along the line-of-sight suggest that only NGC 4438 and M86 are interacting with one another in any substantial way. There is no evidence in either the X-ray imaging or spectral maps for diffuse emission or cool gas bridging between any other pair of galaxies in a manner suggesting galaxy-galaxy interactions. The reported intracluster light (ICL) between M86 and M84 \citep{Rudick2010} therefore appears likely to be a projection effect.  

\subsection{Correlation of the Coolest X-ray Gas and H$\alpha$ Filaments}

\begin{figure*}
\includegraphics[width=0.47\hsize]{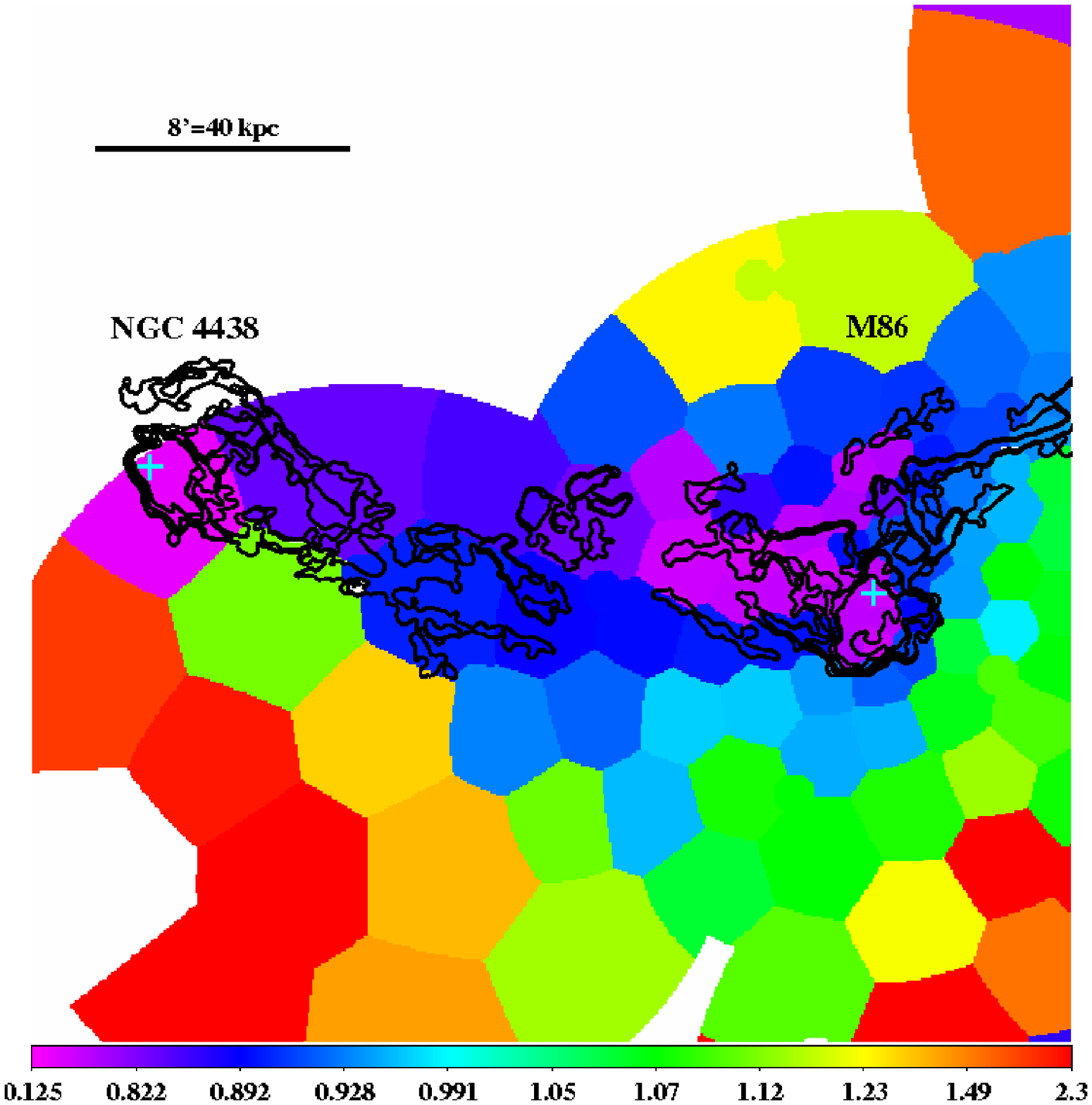}
\includegraphics[width=0.47\hsize]{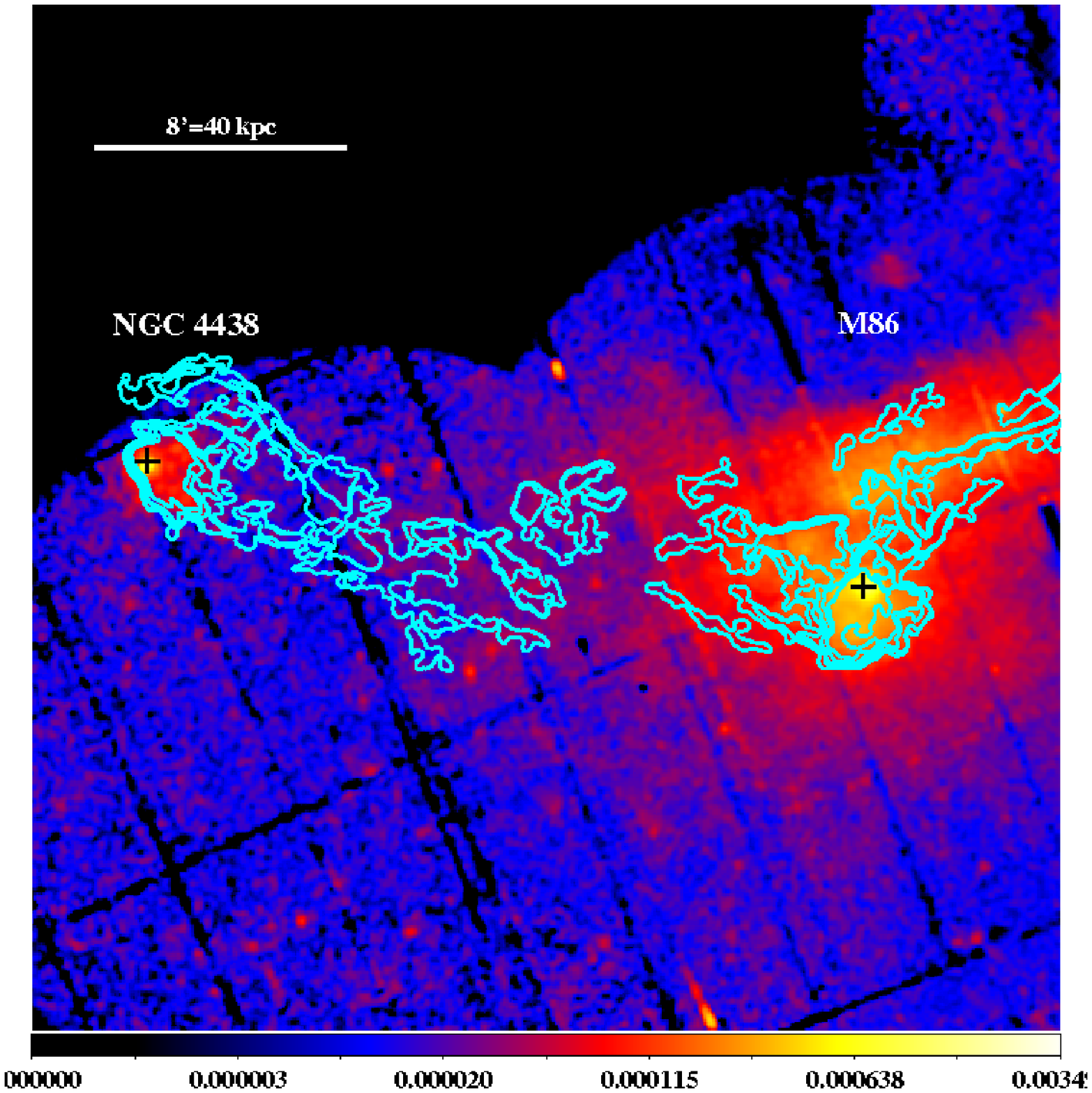}
\caption{\label{WithHA} The correlation between the cool X-ray gas and H$\alpha$ emission detected by Kenney et al. 2008. (a) Temperature map of Figure \ref{Thermo}d zoomed in to emphasize the temperature structure surrounding M86 and overlaid with contours of H$\alpha$ imaging. There is a clear correlation of the H$\alpha$ emission with the coolest gas phases in both the eastern and northern plumes extending out from M86. The centers of M86 and NGC 4438 are denoted by the cyan crosses. (b) The X-ray surface brightness image of Figure \ref{Mosaic}, zoomed in to the same region as (a). The H$\alpha$ image contours are overlaid in cyan. }
\end{figure*}

In Figure \ref{WithHA}a, we show the temperature map of Figure \ref{Thermo}c zoomed in to better show the $\sim 0.6 \keV$ gas plume to the east of M86. We have overlaid the H$\alpha$ filaments identified by \cite{Kenney2008}, which clearly show that the H$\alpha$ filaments are located at the same positions as the lowest temperature phases of X-ray emitting gas. This region, at least the half closer to M86, is also observed to be a site of HI emission \citep{Oosterloo2005}, although based on the velocity structure of that emission it may be primarily associated with NGC 4388 to the south of M86. On the other hand, some of the \ion{H}{i} filaments detected near M86 have velocity structure similar to M86 \citep{Kenney2008}. 

A similar spatial coincidence between cool X-ray emitting gas and H$\alpha$ filaments has been seen in a stripped tail behind the galaxy ESO 137-001 falling into the cluster Abell 3627 \citep{Sun2007}. The stripped gas trailing behind ESO 137-001 is a site of {\it in situ} star formation taking place outside of the galaxy within the ICM. A similar coincidence of cool X-ray emitting gas (at temperatures of $\sim 0.6 \keV$) with filaments of H$\alpha$ emission (at temperatures of $\sim 10^{4} \K$) has also been observed around the BCGs of nearby cool core clusters such as the Perseus Cluster \citep{Sanders2007,Fabian2008,Fabian2011}, Centaurus Cluster \citep{Sanders2002,Sanders2008}, and M87 \citep[][]{Werner2010,Werner2012}.

\subsection{On The Origin of the H$\alpha$ Filaments}

The origin of this $\sim 10^{4} \K$ phase of gas spatially coincident with soft X-ray emission between M86 and NGC 4438 is likely a previous collision between the two. The existing multiwavelength data are all consistent with this scenario, although further data will be required to rigorously confirm this. The X-ray emitting gas in this filament region appears poorer in metals than the gas detected at the center of M86, implying that this cool X-ray emitting gas likely originated in NGC 4438. Measurement biases, due to the presence of multiple temperature phases unaccounted for in our models cannot be ruled out as the reason for the observed low metallicity in this region, however.

In the absence of sources of heating, this stripped gas may be expected to cool to star-forming temperatures, as is observed in the galaxy  ESO 137-001 in Abell 3627 \citep{Sun2007}. Uncertainties in the filament geometry (specifically the depth of the filament regions along the line of sight) limit our ability to determine the cooling time of the cool gas unambiguously. Based on the combined analysis of all of the spectral bins in this filament region discussed above, we can place an upper limit on the integrated cooling rate from $1 \keV$ down to $0.1 \keV$ across the whole region of $\lesssim 0.1 \msolar \yr^{-1}$. This is in agreement with the upper limit on star formation imposed by far UV emission of this same region discussed above. 

Spiral galaxies such as NGC 4438 typically have X-ray halos with an observed temperature of $\sim 0.2 \keV$, while the gas associated with the group halo of M86 has a temperature of $\sim 1 \keV$. The presence of $\sim 0.6-0.7 \keV$ gas in between the two galaxies suggests a possible mixture of gas phases. This ``bridge'' of cool gas between the two galaxies may arise either from ram pressure stripping of NGC 4438 by the surrounding Virgo ICM, or from the recent encounter between NGC 4438 and M86. The stripping medium in this latter scenario could either be the X-ray gas halo of M86 or the intragroup medium surrounding it, both of which could result in the observed filament of $\sim 0.6-0.7 \keV$ gas.

The spatial coincidence between H$\alpha$ emitting filaments with $\sim 0.6 \keV$ X-ray emitting gas is similar to that observed in the centers of cool core clusters undergoing AGN feedback. It is therefore possible that similar physical processes drive the evolution of these distinct filament regions, at least after an initial disturbance. As discussed in \cite[e.g.][]{Fabian2008,Fabian2011,Werner2012}, a potential energy source powering the FIR through UV emission (including the H$\alpha$+\ion{N}{ii} optical line emission) in cooling core clusters are hot ICM particles penetrating into the filaments of cold gas, resulting in collisional ionization and heating \cite[see][]{Ferland2008,Ferland2009}. \cite{Werner2012} propose that shear instabilities due to shocks or the motion of the filaments through the ambient ICM may facilitate mixing between the hot and cold gas phases. The physics of the multiphase gas in the M86 filament region, which is exposed to shear instabilities as it moves through the Virgo ICM and merger shocks due to the infall of the M86 group, may therefore share similarities with the physics of the optical emission line nebulae seen in the cooling cores of galaxy clusters. One key difference between the filament region discussed here with those observed near cool core clusters, however, is the nature of the initial disturbance. Rather than instabilities triggered by AGN feedback, the collision between M86 and NGC 4438 would be the source of the shearing here. Such triggering may be further encouraged by M86's supersonic motion and subsequent shock heating. 

Sites within $\sim 10 \kpc$ of the two galaxies have recently been shown to host CO($1 \rightarrow 0$) line emission, suggesting that molecular hydrogen gas is present between the two galaxies for time scales of at least $\sim 100 $ Myr \citep{Dasyra2012}. This molecular gas is inferred to originate primarily from NGC 4438, similar to what we infer for the X-ray emission. Further observations that place stricter constraints on the star formation rate and detailed spectral properties of this region, as well as the mass of cold gas present from emission lines such as \ion{C}{ii}, will be required to test the connection between the H$\alpha$ emission presented here and that observed in cool core clusters.

\subsection{Combining AGN Feedback with Ram Pressure Stripping}

The detection of a clear tail of X-ray emission trailing to the south of M84 offers another example of a galaxy where feedback and ram pressure stripping work in conjunction to strip its X-ray halo. Ram pressure stripping combined with AGN feedback are also observed in Sersic 159-03 \citep{Werner2011} and the Virgo elliptical galaxy NGC 4552 \citep{Machacek2006,Machacek2006b}, and similar processes are observed at larger scales in the Ophiuchus Cluster \citep{Million2010b} and MACSJ1931.8-2634 \citep{Ehlert2011}. In these systems, the combined influences of these two processes can rapidly suppress or even destroy cool cores associated with the central regions of the galaxy/galaxy cluster. It appears as though a configuration where both AGN feedback and ram pressure stripping occur simultaneously is more common than initially expected.  

The energy injected into the X-ray halo from AGN feedback may allow M84 to be stripped more efficiently now than in its past. At its current stripping rate of $\lesssim 7  \msolar \yr^{-1}$ estimated above, all of the X-ray gas should be stripped from this galaxy within $\sim 200-300 $ Myr, a time scale likely far shorter than the time elapsed since M84 first crossed the Virgo Cluster's virial radius. We determine whether losses from stellar winds may be sufficient to replenish the X-ray halo of M84, assuming a single-age passively evolving stellar population with a Salpeter IMF. We use the stellar mass-loss rate from \cite{Ciotti1991} to estimate this rate, $\mysub{M}{\star}(t)$,  as

\begin{equation}
\mysub{M}{\star}(t) \simeq  1.5 \times 10^{-11}\mysub{L}{B}\mysub{t}{15}^{-1.3}
\end{equation}

\noindent where \mysub{L}{B} is the present-day B-band luminosity in units of \mysub{L}{B,\odot}
\citep[$\mysub{L}{B} = 4.5 \times 10^{10} \mysub{L}{B,\odot}$ for M84;][]{Finoguenov2002} and \mysub{t}{15} is the age of the stellar population in units of 15 Gyr\footnote{This formula is valid in the range from $∼0.5$ to over $15$ Gyr. }. The local stellar mass-loss rate, assuming that the current stellar population is $10$ Gyr old, is of the same order-of-magnitude as the rate of ram pressure stripping ($\mysub{M}{\star} \sim 1 \msolar \yr^{-1}$).

\subsection{On the Frequency of Stripping in Cluster Member Galaxies}
Perhaps surprisingly, each of the four X-ray emitting galaxies detected in this mosaic show evidence of being stripped as they traverse through the ICM. Although these galaxies share some common features, no two galaxies are observed to have stripped gas tails sharing all of the same properties. The prevalence of gas trailing in the wake of galaxies in this busy region of the Virgo Cluster suggests that stripping processes may play an important role for nearly all galaxies in clusters. The particular observational signatures associated with the stripping process may depend strongly on the morphology, mass, and gas content of the host galaxy, but it appears typical that gas initially hosted by galaxies is efficiently stripped by the ICM.

For more massive elliptical galaxies hosting large masses of X-ray emitting gas like M86 (likely the central galaxy of an infalling group) and M84, the stripping is most readily observed as tails of X-ray emitting gas trailing the galaxies. The stripped X-ray gas behind these two galaxies, however, suggest that the astrophysical processes that drive the stripping may vary between different sites. One key difference between the stripped X-ray gas between M86 and M84 is the relative length of the apparent tails, which may indicate differences in the gas physics between the two sites. The tail behind M86 remains roughly uniform in width and coherent for distances of at least $\sim 150 \kpc$ in projection without significant evidence for mixing with the surrounding ICM. The true length of this tail is also likely significantly larger than its observed length in projection due to M86's high velocity with respect to Virgo ($\sim 1500 \km \s^{-1}$) along the line-of-sight. Maintaining such cohesion over these distances suggests that the stripped gas is undergoing a laminar flow even as the galaxy traverses through the ICM supersonically, and subsequently suggests that the stripped gas may have high viscosity. The gas trailing behind M84, on the other hand, is only observable for distances of $\sim 15 \kpc$ in surface brightness and temperature maps, and the lower velocity of M84 with respect to the Virgo ICM along the line-of-sight ($\sim 200-300 \km \s^{-1}$) indicates that this tail is probably not significantly longer than its projected length. The temperature structure in this tail is also observed to increase smoothly with distance from M84. These may be evidence for more efficient mixing between the stripped gas and the surrounding ICM at this site, perhaps associated with a more turbulent flow of the stripped gas. Deeper observations will be required to determine the full length of the stripped gas trailing M84 and subsequently the extent to which turbulence and magnetic fields may be necessary to account for the thermodynamic structure in M84's tail. 

Spiral galaxies contain significant masses of gas at much lower temperatures ($\sim 10-100 \K$). The observed differences in the stripped cold gas behind NGC 4388 and NGC 4438 suggest that the manner in which these phases interact and intersperse with the ICM may vary also significantly from site to site, and that the presence of shear instabilities or shocks may play an important role in these processes. This also appears to be true for the X-ray halos, especially when the disruptions due to the surrounding environment are more extreme. Deeper multiwavelength observations will again be necessary to fully understand the range of gas temperatures present in these stripped tails and test the stripping processes from start to finish.

\section{Conclusions}
We have measured the X-ray spectral properties in a contiguous $\sim 1^{\circ} \times 1^{\circ}$ region spanning four Virgo Cluster galaxies, allowing us to view interactions between them. Our key results are as follows: 

1) There is clear evidence for cool ($\sim 0.6 \keV$) gas trailing to the northwest of M86 as reported in previous studies, to the east of M86 in the direction of NGC 4438, and to the south of M84. M86 appears to be surrounded in a halo of $\sim 1 \keV$ gas, which we interpret as the intragroup medium of a recently merged subcluster undergoing ram pressure stripping by the ambient Virgo ICM. We also observe tentative evidence for stripped X-ray gas to the east of NGC 4388. 

2) The presence of stripped gas associated with all four of the largest galaxies in this field of view indicates that, while the particular astrophysical details may vary from site to site, ram pressure stripping is a common occurrence in cluster member galaxies, and has important implications for the heating and enrichment of the ICM. 

3) Filaments of H$\alpha$ emission surrounding M86 are observed at locations spatially coincident with enhanced X-ray emission at temperatures of $\sim 0.6 \keV$. These filament regions have similar X-ray spectra to H$\alpha$ filaments observed in the centers of cool core clusters. One key difference between this filament region and those detected in cool core clusters, however, is the likely source of the initial disturbance that triggers their formation.  

4) AGN feedback and ram pressure are both observed in M84, and  in tandem these two processes may be encouraging particularly efficient stripping. Without replenishment, M84's X-ray halo gas would be depleted on time scales of $\sim 300$ Myr. Mass loss from stellar winds, however, appears to be sufficient to replenish the M84's X-ray halo, at least in principle.   

5) Our results for M86 and M84 suggest that there is no significant interactions taking place between them. Of these four galaxies, only M86 and NGC 4438 appear to be undergoing any noticeable galaxy-galaxy interactions. 

Disentangling the interactions of the infalling M86 galaxy group in the directions of M84, NGC 4338, and NGC 4438 will require further observations at multiple wavelengths.

\section*{Acknowledgments}
This work was carried out with proposed and archival observations with \xmm, an ESA science mission
with instruments and contributions directly funded by ESA Member States and NASA. This research has also made use of the NASA/ IPAC Infrared Science Archive, which is operated by the Jet Propulsion Laboratory, California Institute of Technology, under contract with the National Aeronautics and Space Administration. Support for this work was provided by the National Aeronautics and 
Space Administration through Einstein Postdoctoral Fellowship
Award Number PF9-00070 (A.S.) issued by the Chandra X-ray 
Observatory Center, which is operated for and on behalf of the National Aeronautics and Space
Administration under contract NAS8-03060. Further support for this work was provided by the 
Department of Energy Grant Number DE-AC02-76SF00515. This work is based [in part] on observations made with Herschel, a
European Space Agency Cornerstone Mission with significant participation by NASA. Support for this work was provided by NASA through award \#1428053 issued by JPL/Caltech.

\bibliographystyle{mnras}
\def \aap {A\&A} 
\def \statisci {Statis. Sci.}
\def \physrep {Phys. Rep.}
\def \pre {Phys.\ Rev.\ E}
\def \sjos {Scand. J. Statis.} 
\def \jrssb {J. Roy. Statist. Soc. B} 


%

\def \araa {ARA\&A}
\def \aaps {A\&AS}
\def \aj {AJ}
 \def \aas {A\&AS}
\def \apj {ApJ}
\def \apjl {ApJL}
\def \apjs {ApJS}
\def \mnras {MNRAS}
\def \nat {Nat}
 \def \pasp {PASP}
\def \gca {Geochim.\ Cosmochim.\ Acta}
\def \prd {Phys.\ Rev.\ D}
\def \prl {Phys.\ Rev.\ Lett.}
\def \ssr {Space Sci.\ Rev.}

\bibliography{M84}

\end{document}